\documentclass[12pt]{article}
\usepackage{authblk}
\usepackage{bm}
\begin {document}

\title {{Nonlocal mechanics of kinetic densities with correlated stresses}}

\author[a,b] {I. {\'E}. Bulyzhenkov}

\affil [a] {Moscow Institute of Physics and Technology,  
    } 

\affil[b] {Lebedev Physics Institute RAS, Moscow, 119991}


\date{}






\maketitle

\begin{abstract}
The non-equilibrium densities of nonlocal mass-energy are self-governed by kinetic stresses toward quasi-equilibrium sub-configurations. System energy integral of continuous matter-extension coordinates its adaptive densities on each hierarchic level of structural sub-organizations. The logarithmic potential for metric fields, kinetic densities, and local stresses in the nonlocal distribution of mass-energy corresponds to the Shannon fundamental limit for information rates. The phenomenological pulls of Newtonian gravity can be quantitatively replaced by volume kinetic pushes from local stresses of continuous space-matter with moving inertial densities. Thus, the observable phenomenon of gravitation between volumetric parts of distributed mass-energy is interpreted in this Cartesian nonlocality  of self-governed material space through correlated accelerations of continuous kinetic densities with local inertia. The gravitational palliative of negative potentials is replaced with the monistic approach to Cartesian matter through always positive kinetic mass-energy.
\end{abstract}


\section{Introduction}

Since 1914, Einstein used the famous geodesic  law  $c^2 u^\nu\nabla_\nu u_{\mu} = 0$ for the free fall of the probe body, which does not contribute to the metric fields in the covariant derivatives $\nabla_\nu u_\mu \equiv \partial_\nu u_\mu -\Gamma_{\nu \mu}^\rho u_\rho.$  This advanced modeling of 4-accelerations in external fields formally supports Newton\rq{s} gravity of point masses and, therefore, their \lq\lq divine action-at-a-distance\rq\rq{}. But why does the probe mass not affect locally the metric field exerted to \lq{}independent\rq{} matter of unadaptive probe densities? The original version of Einstein\rq{s} metric fields \cite {Ein} was unavoidable refereed in 1916 by the Newton gravitational theory for localized matter (allegedly observed in  local measurements) in external fields of weak mutual attractions.  In this historical way, negative gravitational energies were imperceptibly accepted by modern physics,  which ultimately led the point matter model to the black hole, dark energy, and other quite objectionable
 notions.  

 In order to restart mechanics of visible (macroscopic) bodies and the 1st fundamental force of interactions from the very beginning, we propose mathematically restarting the theory of inertial motion  from 1644  \lq\lq Philosophiae Naturalis\rq\rq{} of Descartes, and not from 1687 \lq\lq Philosophiae Naturalis Principia Mathematica\rq\rq{} of Newton. The relativistic laws of General Relativity (GR) can be in good agreement with the Cartesian ontology of continuously distributed masses. And continuous electric densities are also inherent in Maxwell\rq{s} equations due to their exact analytical solutions \cite {Bul}. In general, the Cartesian matter-extension of continuous mass-energy without void regions \cite{G} suggests a non-equilibrium distribution of material densities in the multi-vertex kinetic system. The visible motion of the mass-energy vertexes may not follow the  Newtonian action-at-a-distance, but the kinetic self-organization of the mechanical system through local pushes of material (positive) densities in their non-local unity under the constant energy integral. The non-equilibrium motion of dense areas next to traceable vertexes may better correspond to the universal tendency of a closed system to its dynamic quasi-equilibrium or oscillations around the equipartition of kinetic chaos and kinetic ordering \cite{B}, and not to the Newtonian evolution in negative gravitational potentials.

The negative gravitational energy or potential is a palliative phenomenology to model interactions in a  multi-body systems. But negative energies should not exist in the kinetic reality of measurable (positive) energies. And extra (potential) fractions of energy should not be employed in the 1644 monistic approach to nature through the endless kinetics of inertial densities with local pushes due to mechanical stresses in continuous mass-energy. In other words, the Cartesian worldview monistically associates physical reality only with positive kinetic energies, that is, without the notion of gravitational fields. In this article, we will return to such monistic metaphysics of relativistic kinetic (positive) energies and will reject mathematically the Newtonian gravitational attraction by negative (nonexistent) energies. Distant attractions will be replaced by local self-accelerations of positive asymmetric densities formed by a nonequilibrium metric potential in a multi-vertex system of kinetic energies. 

The Cartesian approach can draw a lot from General Relativity due to the amazing finding of Kuhn  \lq\lq Einstein\rq{s} theory can be accepted only with the recognition that Newton\rq{s} was wrong\rq\rq{} \cite {Kuh}. Unlike Newton,  Einstein\rq{s} metric energies never relied on negative values. Special Relativity (SR) introduced positive (kinetic) rest-energy $mc^2$  and allowed GR to work only with positive kinetic densities $\mu c^2 {\sqrt {{g_{oo}} / 1-\beta^2}} > 0$ of real bodies at any metric field $\sqrt {g_{oo}} $.  GR does not work with negative mechanical energies and must revise Newtonian references with independent concepts for kinetic and gravitational energies. Gravity cannot be postulated as an independent phenomenon, because only kinetic energies determine the geodesic acceleration or deceleration of matter in the relativistic organization of metric energies \cite {B}. The goal of our kinetic alternative to Newtonian gravitation is to develop suitable mathematics for  Descartes\rq{} monism and to propose the non-local world hierarchy of stable non-equilibrium sub-systems  for further use in physics, biosciences, informatics, and engineering.

\section {Logarithmic potential of continuous space-matter and Shannon\rq{s} information potential}

Static organization of Cartesian material space $ {\bf x} \equiv  {\bf r} $ can form radially  distributed mass-energy  $mc^2 = c^2\int  \mu (r)  4\pi r^2 dr$ with the equilibrium metric densities 
$ \mu c^2 = mc^2 r_o/4\pi r^2 (r+r_o)^2  = \partial_i 
({ \sqrt {g_{oo} }}  g^{oo} \partial^j { \sqrt {g_{oo} }}  )   =   - \partial_i {D^i}(r) $ \cite {Bul,BulG}.  These inertial mass-energy densities are defined by the post-Newton metric field ${\hat {\bf r}}D(r) = - {\hat {\bf r}}\partial_r W(r) = - {\sqrt G} m {\hat {\bf r}}/r(r+r_o)$ or the metric potential for strong interaction fields $W{(r)} =  {\varphi_{\!\!_{_G}}} ln { \sqrt {g_{oo}(r) }}= - {\varphi_{\!\!_{_G}}}ln [1+ ({\sqrt G} m /r{\varphi_{\!\!_{_G}}} ) ] $ in Euclidean 3-space, where $(g_{oi}g_{oj}/g_{oo}) -  g_{ij}  = \delta_{ij}$,  $g_{oo} = 1/ g^{oo} = r^2/(r+r_o)^2$, and $r_o \equiv Gm/c^2 > 0$.
 The metric potential $W (g_{oo})$ in continuous space-matter corresponds to the Shannon maximal potential   \cite {Sha,Shan} for incoming / outgoing information rates  $C (N) =
\pm w\ \! log_{_2} \left[ 1+ ({P}/ {N w}) \right]$. This analogy clarifies from $P/wN \Rightarrow  {\sqrt G} m /r{\varphi_{\!\!_{_G}}}$ the mechanical charge self-potential $\varphi_{\!\!_{_G}}\equiv c^2/ {\sqrt G}$ as the energy-information bandwidth $w$. The spatial path $r$ corresponds to the noise valuer $N$. Therefore, large spatial distances  prevent high  transmission rates for energy-information exchanges between material densities. 

A static multi-vertex system of metric fields and their mass-energy densities can also  be described \cite {BulG,Buly} with the logarithmic  potential $W_{sys}(x)$ $ = - ln (1+ \sum^n_{1} Gm_k/R_k  )$, which   generalizes  the Shannon information limit for multiple sources $P_k$:
\begin {eqnarray}
\cases { 
C_{sys} = \pm w\ \! log_{_2} \left ( 1+  \sum^n_{1} \frac {P_k}{w N_k } \right ), \cr
W({\bf x}) = - {\varphi_{\!\!_{_G}}} ln \left ( 1+ \sum^n_{1} \frac {m_k{\sqrt G}} { {\varphi_{\!\!_{_G}}}|{\bf x}-{\bf a}_k|}  
\right ). }  
\end {eqnarray}
To understand the mass-energy organization according to the Shannon information law (1) for a multi-source system, we first analyze  the static distribution of elementary energies $m_k c^2 \equiv {\sqrt G}m_k \varphi_{\!\!_{_G}} \equiv  r_k {\varphi^2_{\!\!_{_G}}}$, centered around the vertexes ${\mathbf a}_k$ in 3D material space:      
\begin{equation}
{\mathbf D}_{sys}(\mathbf{x}) \equiv - {\bm \nabla } W_{sys} (\mathbf{x}) = \sum^n_{k=1}\frac{(-\varphi_{\!\!_G}r_k )}
{\left(1+\sum^n_{1}\frac{r_p}{|\mathbf{x-a}_p|} \right)}\frac{(\mathbf{x-a}_k) }{\left|\mathbf{x-a}_k\right|^3} 
\equiv  \sum^n_{k=1} {\mathbf D}_{k} (\mathbf{x}) .
\end{equation}
Here $\mathbf{D}_k \equiv ({\mathbf a}_k -{\mathbf x}) {\sqrt G} m_k / {|\mathbf{x-a}_k|^3} [1+\sum^n_{1}({r_p}/{|\mathbf{x-a}_p|)}  ]$ is an elementary asymmetrical field which, unlike the radial fields of Newton/Coulomb point particles, depends from   positions of all system vertexes $\mathbf{a}_p$ (in the common denominator). 
  \lq Elementary\rq{} sub-densities ${\sqrt G}\mu_{k}$ of the system charge density,  ${\sqrt G}\mu_{sys} \equiv {\mathbf D}^2_{sys}/4
\pi\varphi_{\!\!_G} 
 \equiv - div  \mathbf{D}_{sys}/4\pi \varphi_{\!\!_G}$ $  \equiv - \sum_1^n div \mathbf{D}_{k}/4\pi\varphi_{\!\!_G}  \equiv {\sqrt G}\sum_1^n \mu_{k}$,  also depend on all vertexes
 ${\mathbf a}_p$. 
Now,  one can calculate from (2) the  continuous mass density $\mu_{sys}c^2$ and the relativistic rest-energy, $\sum_1^n m_k(t)c^2 = const$,
 of a closed non-equilibrium system of quasi-equilibrium elements. These elements are not closed formations and  they can contribute to the steady system as with constant  \lq elementary\rq{} energies $m_k(t)c^2 = C_k$, as well as with time-varying energy contributions
($m_k(t)c^2 = C_k - \hbar \omega_{kl}(t)$ and $m_l(t)c^2 = C_l + \hbar \omega_{kl}(t)$
 due to mutual inelastic exchanges within the closed system): 
\begin{equation}
\cases { 
{\varphi_{\!\!_G}}\!{\sqrt G}\mu_{sys}\! (\mathbf x) \! 
\equiv\! \frac { \left (\! \sum_1^n {\mathbf D}_k\! \right)^2   } {4\pi} 
\!\equiv \!\frac { \left (\!\sum_1^n\! \frac{Gm_k^2}{\left|\mathbf{x\!-\!a_k}\right|^4}\!\right ) \! + \!
 \sum^{n}_{1} \! \frac{ (\mathbf{a_k \!-\! x}){\sqrt G} m_k}{\left|\mathbf{x\!-\!a_k}\right|^3} \left (\sum^{n-1}_{s\neq k}\frac{ (\mathbf{a_s\!-\!x}) {\sqrt G}m_s}{\left|\mathbf{x\!-\!a_s}\right|^3} \right ) 
   }
{4\pi \left(1+\sum^n_{1}\frac{r_p}{|\mathbf{x-a}_p|} \right)^2},
\cr \cr
\int\! c^2\!\mu_{sys} ({\mathbf x}) d^3\!x \equiv \int \!\frac {d^3\!x            }{4\pi}  {\mathbf D}^2_{sys}
\equiv  - \int \!\frac {d^3\!x}{4\pi} \partial_i \left (\sum_1^n {\mathbf D}^i_{k}\right )\equiv
 \sum^n_1\!m_k c^2 = const,
\cr\cr
\int\!\! c^2\!\mu_k d^3\!x \!\equiv \! \int\! \!\frac {d^3\!x   {\mathbf D}_k {\mathbf D}_{sys}        }{4\pi}  
\!\equiv\!  { \int}\!\! \frac  {d^3\!x   \frac { (\!\mathbf{a_k\!-\!x}\!){\sqrt G}\! m_k} {\left|\mathbf{x\!-\!a_k}\right|^3}  \! \!\left (\!\sum^n_1\!\frac{ (\mathbf{a_s-x}){\sqrt G} m_s}{\left|\mathbf{x-a_s}\right|^3}\!\! \right ) 
   }
{4\pi \left(1+\sum^n_{1}\frac{r_p}{|\mathbf{x-a}_p|} \right)^2}\!
\equiv \!
 - \!\int \!\!\frac {d^3\!x}{4\pi} \partial_i D^i_k\! \equiv\!
 m_k c^2 .}
\end{equation}
The energy density of the system, $ \mu_{sys}c^2 \equiv (\! \sum_1^n {\mathbf D}_k\! )^2/4\pi   $, is positive at all points of space-matter. There are no negative gravitational energies in the kinetic reality of non-locally distributed integral of the system mass density $\mu_{sys}$ with $\sum_1^n m_kc^2 = const $.

The numerator of the system energy density ${\varphi_{\!\!_G}}{\sqrt G}\mu_{sys} = c^2\mu_{sys}$ in (3) has two sums,  $(\sum_1^n\{...\})$ and $\sum_1^n\{... (\sum^{n-1}_{s\neq k}\{...\}) \}$, of  mono-radial and bi-radial functions, respectively. Such different topologies of \lq elementary\rq{} inertial contributions  mean that the elements in the system acquire new physical properties, and the system is always more complex than the sum of isolated elements. The integrated energy in the last line of (3)  takes into account the mono-radial part of the \lq elementary\rq{} density $\mu_kc^2$ and its interference contribution to the system depending on the specific configurations of other elements. The integral sum $m_k c^2$ does not depend on positions of other vertexes $\mathbf{a_s}$ in a closed system. Consequently, exchanges of radiation and wave information between quasi-equilibrium organizations of elementary energies can occur independently from distances and the 1st force between observed vertexes. Exchanges by quantized mechanical waves (or kinetic photons of inertia) are independent on  distances and  require common resonance properties or similarities of overlapping receivers and transmitters in a shared material space.

\section {Local self-pushes of non-equilibrium densities in non-local systems}
The static mono-vertex solution $g_{oo} = r^2/(r+r_o)^2$ for the consistent mass-energy densities $\mu c^2$ in (3) allows the steady coexistence of  very dense and very rare regions of the continuous non-local distribution of the energy integral $mc^2 \equiv r_o \varphi^2_{\!\!_G}$.  In the developed metric approach to kinetic (measurable) energy, the quasi-equilibrium organization of the elementary integrals  $m_kc^2$ can participate in non-equilibrium dynamics of mechanical densities of each quasi-equilibrium metric system at a higher level of the world energy hierarchy. Such a multi-level hierarchy of almost closed systems with inside non-equilibrium dynamics allows a high concentration of collective inertia, energy and information in some 3D regions  of space-matter and their invisible rarefaction in others. Continuous non-equilibrium densities of almost constant energy integrals of quasi-equilibrium sub-systems have adaptive properties of self-governed distributions  to maintain time-averaged patterns of spatial densities. Thus, energy-information nonlocal control is inherent not only to vital matter but also to inert one.

Let us look at a two-vertex system of  distributed mass-energies $m_1c^2 \equiv  
\int  d^3 x ({\mathbf D}_1^2 + {\mathbf D}_1 {\mathbf D}_2) /4\pi =  \varphi^2_{\!\!_G} r_1 $ and  $m_2c^2 \equiv  \int  d^3 x ({\mathbf D}_2^2 + {\mathbf D}_2 {\mathbf D}_1)/4\pi =  \varphi^2_{\!\!_G} r_2 $  prior to discuss fundamentals of non-local coordination in more complex mechanical organizations. The elementary mass-energy integrals in the binary system are related  to both  vertexes  $\mathbf{a}_1$ and $\mathbf{a}_2$. But why it is possible to   measure the  force of attraction  between  the vertexes, $\propto L^{-2} = |\mathbf{a}_2-\mathbf{a}_1|^{-2} $,  if one requires the constancy, $m_{1/2}c^2= const$, of quasi-equilibrium elementary energies during  their geodesic falls in the absence of radiation exchanges or inelastic frictions? 
In other words, why and how do the centers of extended masses exhibit the Newtonian action-at-a-distance of alleged pulls if the energy-information space of positive kinetic densities does not have negative mechanical energies at all?

Again, the material space of non-local energy can rely in Cartesian mechanics only on positive kinetic densities forming everywhere the continuous  distribution of consistent inertial densities $ \mu (x^\prime) > 0 $.  Such space-matter has inhomogeneous stresses, and we use  the Coulomb-Lorentz  3-force ${\sqrt G}\mu_{sys} {\bf D}_{sys}({\bf x})$ due to the fundamental analogy of (2) with vector electrodynamics.  Integrating these force densities over the ultra-small volume   $\Omega_1 \approx 4\pi d^3 /3 $, $r_1 \leq d \ll L$ around the static vertex of the first partner,   
  \begin{eqnarray}
F_{1}^x\! = \!\!\int_{\Omega_1} \!\!\! \!{\sqrt G}\mu_{sys}  {D}^x_{sys} dxdydz \!
= \!\int_{\Omega_1} \!\frac {d^3x} {4\pi\varphi_{\!\!_G}}  ({\mathbf{D}_1^2 + 2 \mathbf{D}_1  \mathbf{D}_2} +\mathbf{D}_2^2 )
(D^x_1 + D^x_2)  
\cr\! \approx \! 
\!\!\int_{\Omega_1}\! \!\!\frac {d^3x} {4\pi\varphi_{\!\!_G}} 
 (\mathbf{D}_1^2 D^x_1\! +\! \mathbf{D}_1^2 D^x_2  \!+\! 2 \mathbf{D}_1  \mathbf{D}_2 D^x_1\!)\! = \!\frac {\varphi^2_{\!\!_G} r_1r_2} {L^2} \left (\!\frac {1}{6}\!+\! \frac {1}{2} \!+\! \frac {1}{3}\!\right)\!  = \!
 \!\frac {Gm_1 m_2} {L^2},
\end{eqnarray}           
it is possible to calculate  the integral force $F_{1}^x $, applied to  the ultra-small volume $\Omega_1$  towards the second vertex in ${\bf a}_2 = \{L;0;0\}$, with $r_1, r_2 \ll L$. 
The similar integration of the system densities ${\sqrt G}\mu_{sys} {\bf D}_{sys} $ leads from (2)-(3) to the opposite force,  ${\hat {\bf x}}F_{2}^x = - {\hat {\bf x}}F_{1}^x = - {\hat {\bf x}} Gm_1m_2/L^2$, applied to an  ultra-small  volume around the second vertex in  Cartesian material continuum.

There is no \lq\lq divine action-at-a-distance\rq\rq{} in the closed system of directly contacting Cartesian partners with the shared energy density. And there are no negative gravitational energies behind the inhomogeneous distribution of the mass-energy density $\mu_{sys}({\bf x},{\bf a}_1, {\bf a}_2 )c^2 > 0$ in  continuous space-matter   with the  logarithmic potentials (1) to transmit  information and energy. The energy density in the system gains interference fractions, ${\mathbf D }_1( {\bf x}, {\bf a}_1)  
{\mathbf D}_2 ( {\bf x}, {\bf a}_1)/2\pi$. This interference is disappearing for distantly isolated  particles with steady  radial densities
$\mu_{1/2} c^2 = \varphi^2_{\!\!_G} r^2_{1/2}/4\pi ({\bf x}-{\bf a}_{1/2})^2 (|{\bf x}-{\bf a}_{1/2}|+r_{1/2})^2$. Again,  mathematics  (3) and (4) claims that the energy-information system is not an ordinary sum of individual elements, but always acquires new physical properties after their join gathering. Due to the common denominator, the system field (2) gains nonequilibrium asymmetry near the both vertexes of former radial elements. This system asymmetry of the metric field ${\bf D}_{sys}$  in (4) leads to two non-equilibrium self-forces next to the equal force density ${\mathbf D}_1^2 ({\bf x})  D^x_2 ({\bf x})$. The most intense information and energy transfers occur in dense volumes near the elementary vertex.  Despite the slow nonequilibrium self-assembling of a large nonlocal system, its continuous elements can more quickly assemble into quasi-equilibrium nonlocal subsystems. Dense quasi-equilibrium regions of such subsystems can be separated by spatial distances of space-matter with an invisible low, but always finite density everywhere. Limited human sensitivity forms the illusion of the visible separation of quasi-equilibrium dense areas, say, spatially separated apples on a tree and the ground.

 The asymmetry of the field densities and the static   force (4) determines the local acceleration of dense regions in the nonlocal energy organization. Not distant gravitational pulls, but local inertial pushes cause the apple to fall to the ground in their nonlocal organization of joint kinetic energy. The metric distribution of material densities in the system with the basic potential (1) describes the non-equilibrium asymmetry of Lorentz forces around the vertexes. By placing one selected vertex at the origin, ${\bf a}_1 = 0$, we can calculate  the pushing self-action $F_{1}^i\! = \!\int_{\Omega_1} \! d^3x\!{\sqrt G}\mu_{sys} {D}^i_{sys} $ of the ultra-small sphere $\Omega_1 = 4\pi d^3/3$, 
$r_1 \leq d \ll |{\bf L}_s| \equiv |{\bf a}_s - {\bf a}_1|$, which originates from the asymmetry of the Coulomb-Lorentz force densities, 
 \begin{eqnarray}
F_{1}^i\!\! =\!\!
 \int_{\Omega_1}\!\!\!\!\!d^3x\!  \frac { \!\!\sum_1^n \!\!\frac{G m_k^2}{\left|\mathbf{x-a_k}\right|^4}\!\!+ \!
 \!\sum^{n}_{1} \!\! \frac{ (\!{\mathbf a_k\! -\! \mathbf x}\!){\sqrt G} m_k}{\left|\mathbf{x-a_k}\right|^3}\!\! \left (\!\sum^{n-1}_{s\neq k}\!\!\frac{ (\mathbf{x-a_s}\!){\sqrt G} m_s}{\left|{\mathbf a_s-\mathbf x}\right|^3} \!\right ) 
   }
{ 4\pi\varphi_{\!\!_G} \left(1+\sum^n_{1}\frac{r_p}{|\mathbf{x-a}_p|} \right)^3} \!\!
  \sum^n_1\!\! \frac{ (\!a^i_k\! -\! x^i)\!{\sqrt G} m_k} {\left|\mathbf{x-a_k}\right|^3} \cr
		\cr
	 \!\approx\! \!\int_{\Omega_1} \!\frac {d^3x} {4\pi\varphi_{\!\!_G} }\! \!\left (\!\mathbf{D}_1^2\! \sum_1^n\! D^i_k \!   + \!2 D^i_1  \mathbf{D}_1 \!\sum^{n-1}_{s\neq 1}\! \mathbf{D}_s \! \right )\!    
 =\!
	- \!\int_{\Omega_1}\! 
		 \frac {d^3x  \varphi^2_{\!\!_G} r^3_1 x^ i  }   
{ { 4\pi\!\left|\mathbf{x}\right|^7}\!\left(\!1\!+ \!\sum^n_1\frac{r_p}{|\mathbf {x-L}_{p}|}\! \right)^3}  \!
   \cr\cr
	\!+\!\int_{\Omega_1} 
		\! \frac {d^3x \varphi^2_{\!\!_G} r^2_1
\!\sum^{n-1}_{s\neq 1}\!\frac{ ({L^i_{s} - x^ i}) r_s } {\left|\mathbf{x-L_{s}}\right|^3}  
   }
{ { 4\pi\left|\mathbf{x}\right|^4}\left(1+ \sum^n_1\frac{r_p}{|\mathbf {x-L}_{p}|} \right)^3} \!
+\! \int_{\Omega_1}\! 
		\! \frac { d^3x
  { {\varphi^2_{\!\!_G} r^2_1 x^i} }
   \left (\!\sum^{n-1}_{s\neq 1}\frac{ { \!\mathbf{x}\!}({\mathbf L_{s} - \mathbf x}  \!)r_s}{\left|\mathbf{x-L_{s}}\right|^3} \!\right ) 
   }
{2\pi\! \left|\mathbf{x}\right|^6\!\left(\!1\!+\! \frac {r_1}{|\bf x|\!}+\!\sum^n_{p\neq 1}\frac{r_p}{|\mathbf {x-L}_{p}|}\! \right)^3} 
  \cr
	\!\!\approx - \!\int_{\Omega_1}\!   \frac {{d^3x}  \varphi^2_{\!\!_G} r^3_1 x^i } {4\pi\!\left|\mathbf{x}\right|^7\! \left (\!1 + \frac {r_1}{|\bf x|}\!\right )^3 }  
	 	\left(
	1\!+ 	\!  \sum^{n-1}_{s \neq 1}\frac{r_s} 
	{	\left (1 + \frac {r_1}{|\bf x|} \right ) L_{s} {\sqrt  { 1-2 {\frac { {\mathbf x \mathbf L }_{s} }{ L^2_{s}}} +  \frac{x^2}
	{L^2_{s}}   }}  
	} 
	\right)^{-3}
	\cr
	+\!\int_{\Omega_1}\! \!\! {d^3x} 
		\! \frac { \!\varphi^2_{\!\!_G} r^2_1
\!\sum^{n-1}_{s\neq 1}\!\frac{ (L^i_{s}-x^i) r_s } {\left|\mathbf L_{s}\right|^3}  
   }
{ { 4\pi\left|\mathbf{x}\right|^4}\left(1+    \frac {r_1}{|\mathbf x|} \right)^3} \!
	+\! \int_{\Omega_1}\! \! \!{d^3x} 
		\! \frac { 
  { {\varphi^2_{\!\!_G} r^2_1 x^i} }
\sum^{n-1}_{s\neq 1}\frac{ ({\mathbf x} {\mathbf L}_s - x^2) r_s}{\left|\mathbf L_{s1}\right|^3} 
   }
{2\pi\! \left|\mathbf{x}\right|^6\!\left(\!1\!+\! \frac {r_1}{|\bf x|\!}  \right)^3}
\cr \!= \!\!\sum^{n-1}_{s\neq 1}\!\!\int_{\Omega_1}\! \!\! 
 \frac { {d^3x} \varphi^2_{\!\!_G} r^2_1 r_s  } {4\pi\!\left|\mathbf{x}\right|^4\! \!\left (\!1\! + \!\frac {r_1}{|\bf x|}\!\right )^3\!\!L^3_s }\!
\!\left (\! 
 \frac {3x^i r_1 ({\mathbf x} {\mathbf L}_s) }  { |\mathbf{x}|^3\!\! \left (\!1\! + \!\frac {r_1}{|\bf x|}\!\right ) } 
 \!+\! L^i_s\! +\! \frac {2x^i ({\mathbf x} {\mathbf L}_s)}{|\mathbf{x}|^2} \! \! \right)\!\!=\! \!\sum^{n-1}_{s\neq 1}\!\! \frac {G\!m_1\!m_s {L}^i_s}{L^3_s} 
  				\end{eqnarray}

At first glance, we simply derive the well-known Newtonian pulls between distant vertexes. But our integration over the localized volume around the origin represents exclusively local self-pushes of asymmetric kinetic densities to all other vertexes in the non-local metric organization. In such Cartesian space-matter there are no negative energies and distant gravitational pulls.  But there are adaptive self-pushes of asymmetric kinetic densities toward their non-equilibrium partners. 
 As an elastic property of one non-local system with constant elementary energies, $m_kc^2 = const$ at all time instants, non-equilibrium self-pushes (5) occur simultaneously throughout the whole material organization. Despite the formal dependence of these elastic interactions on distances in (5), such coordinated forces do not have retardation in metric space-time. The internal transfer of radiation energy between extended elements (when $m_kc^2 \neq const$ and $\sum m_kc^2 = const$) does not disappear on the spatial distances, but these inelastic exchanges depend on retarded/advanced time \lq distances\rq{}  between elementary vertexes. Wave exchanges can demonstrate both retarded and advanced properties in one non-locally organized continuum with constant energy, but always retarded properties in dissipative exchanges between independent systems with separated energy balances.   
 
  \section {Conclusion} 

 Based on the minute particles streaming through space, Nicolas Fatio de Duillier from the Republic of Geneva first criticized Newtonian gravity using local kinetic pushes in the 1690 letter to Huygens. Since 1748, this collision-based approach has become known as the kinetic gravitation of Georges-Louis Le Sage, another mathematician from Geneva.  However, mechanical collisions with invisible particles should warm and slow down celestial bodies,  which contradicts the available observations. Nonetheless, in 1873, Lord Kelvin tried to improve the Le Sage kinetic theory, appreciated also by Maxwell  as \lq\lq ... seems to be a path leading towards an explanation of the law of gravitation, which, if it can be shown to be in other respects consistent with facts, may turn out to be a royal road into the very arcana of science\rq\rq {} \cite {Max}.

In 1742 Mikhail Lomonosov proposed to describe mutual  attractions of visible bodies by the local pressure of invisible liquid-matter \cite {Lom,Lomo}.  
The above derived  self-accelerations by local mechanical stresses can reinforce Lomonosov\rq{s} super-penetrating liquid-matter through instant fields in volumetric forces (5) and can maintain the ultra-mundane corpuscles of Le Sage for inelastic exchanges in the energy balances (3).
More general, the local relations (2) - (5) for mechanical densities of space-matter can shed new light  onto the nonlocal grounds for inertial mass-energy organization and for nonlocal communications within inert and living matter. Lomonosov applied his alternative model with the local mechanical pressure of invisible matter-liquid  (associated with Plato\rq{s} ether) to detail the sequential transmission of a nerve impulse for sensitivity, smell and vision in the paper \lq\lq Word on the Origin of Light\rq\rq{.} If critics of Newtonian gravitation in Switzerland and Russia were right indeed regarding local pushing for any mechanical acceleration, the 1st fundamental force  of \lq nonexistent gravitation\rq{} between distant bodies can be locally controlled. In particular,  telekinesis of macroscopic material volumes can be quantitatively related to the laboratory induced asymmetry of density integrands in (5).   

Retarded/advanced relationships between localized charges and waves were introduced for energy-information exchanges between  independent systems of energy balance,  but not for one non-local organization with locally bound field and current densities, as in the Maxwell equations. The elastic forces  (5) and all their components are instantaneously agreed with all asymmetrical densities over the whole nonlocal system of constant mass-energy.   
The dissipation-less dynamics of elements can take place under the constancy of elementary energies in (3), $m_k c^2 = const$. The corresponding conservation  $m_k = const$ for all and each inertial elements is not related to advanced/retarded exchanges of wave signals.  Such a system coordination of instant inertial fields, elementary kinetic densities and local tensions in the non-equilibrium organization of mass-energy distributions explains the Laplace paradox of the Solar system stability over the calculated limit of several centuries \cite {Lap}. 
Again, the instant geodesic self-governance of kinetic energies occurs without radiation exchanges. Wave energy exchanges  slowly disrupt the geodesic orbits of planets in the temporal processes of dissipation, which are in parallel to the elastic oscillations over the Kepler orbits. New experimental schemes can be re-invented to study Cartesian matter-extension with retarded, advanced and instant nonlocal correlations like in the telescope probes of Nikolai Kozyrev \cite{Koz}  and the Baikal Deep-Sea Experiment on macroscopic entanglement and forecast of the solar activity \cite {Kor}.

The asymmetric integrands in the kinetic integrals (5) propose to revisit the \lq gravitational\rq{} tidal waves by carefully integrating the kinetic self-pushes over multi-vertex macroscopic volumes $ \Omega $ on the sea surface in the nonlocally correlated densities of the Earth-Moon-Sun system. The self-governed space-matter can reveal unexpected anomalies of the Newton model in the \lq gravitational\rq{} landing on small asteroids and comets due to the non-locally organized flows of kinetic (positive) mass-energy in the multi-vertex material continuum of inertia.  There are many applied consequences from the \lq obsolete\rq{} pressure of liquid-matter of Lomonosov and drag collisions of Fatio - Le Sage in the under-appreciated monistic view on kinetic energies. In conclusion, the proposed reading of correlated kinetic stresses in continuous space-matter can be useful for a better understanding of the nonlocal self-organization of visible inert and living substances in dense inertial regions of the nonlocal Universe.




	\end {document}